# Lack of Fusion in Additive Manufacturing: Defect or Asset?

Jenniffer Bustillos[a,b], Jinyeon Kim[a], Atieh Moridi[a,b,*]

[a] Sibley School of Mechanical and Aerospace Engineering
Cornell University
124 Hoy Road, 469 Upson Hall
Ithaca, NY 14853, USA

[b]Kavli Institute at Cornell for Nanoscale Science
420 Physical Sciences Building
Ithaca, NY 14853, USA

*Correspondence to: moridi@cornell.edu

**Abstract**
Rapid cooling rates and stochastic interactions between the heat source and feedstock in additive manufacturing (AM) result in strong anisotropy and process-induced defects deteriorating the tensile ductility and fatigue resistance of printed parts. We show that by deliberately introducing a high density of lack of fusion (LoF) defects, a processing regime that has been avoided so far, followed by pressure assisted heat treatment, we can print Ti-6Al-4V with reduced texture and exceptional properties surpassing that of wrought, cast, forged, annealed, and solution-treated and aged counterparts. Such improvement is achieved through the formation of low aspect ratio α-grains around LoF defects upon healing, surrounded by α-laths. This occurrence is attributed to surface energy reduction and recrystallization events taking place during healing of LoF defects. Our approach to design duplex microstructures is applicable to a wide range of AM processes and alloys and can be used in the design of damage tolerant microstructures.

## 1. Introduction

Additive manufacturing (AM) has emerged as a revolutionary technique to make complex metallic parts with high strength, considerable weight savings and without the design constrains of traditional manufacturing processes [1,2]. However, AM parts are characterized with inferior ductility and fatigue resistance as compared to conventionally processed alloys affecting the widespread adoption of the technology [1,3,4]. In the case of AM Ti-6Al-4V, rapid cooling rates during the manufacturing process result in formation of metastable α' martensitic phases which has superior strength as compared to that of stable α, and α+β phases[5–8]. However, the stress incompatibility between residual β-phase and α' martensite, high dislocation density in α', coarse columnar β-grains, and process-induced defects are responsible for the reduced ductility and strong texture in printed parts [1–12]. Intrinsic thermal cycles during consecutive layer deposition, and low thermal conductivity of Ti-6Al-4V (7 W·m·K$^{-1}$) facilitate the accumulation of heat leading to directional thermal gradients that favor columnar grain growth [4,13]. In addition, inevitable porosities found in AM parts are largely attributed to the transfer of trapped atomization gas porosity in feedstock powders, melt pool instabilities, underdeveloped melt pools, and the vaporization of elements [1,3,4,11,12,14,15].

Several strategies have been developed to overcome the above-mentioned challenges limiting the performance of as-printed alloys. To control the solidification microstructure, high-intensity acoustic vibration has been adopted to achieve columnar to equiaxed transition of prior β-grains via cavitation-assisted nucleation [16]. The addition of solute elements such as Cu, B, and Si to the Ti-6Al-4V have enabled a similar columnar to equiaxed transition by increasing the constitutional supercooling within the melt [11,17–20]. However, these methods are not universally applicable to different AM techniques or a wide range of materials. For example, acoustic vibrations cannot be applied to powder bed fusion (PBF)-AM as it disturbs the powder bed. In addition, finding an effective nucleant remains challenging for several alloys and it changes the composition of the material which might not be desirable for certain applications.

Hot Isostatic Pressing (HIP) has been successfully used to eliminate pores in printed parts and is now emerging as an important post processing step to improve their mechanical properties (i.e. ductility and fatigue resistance)[21]. The simultaneous application of heat and pressure to as-printed parts leads to the collapse of pores via plastic deformation-aided diffusion[22]. HIP is now routinely performed on printed samples used for fatigue-critical applications to shift the defect-initiated failure mode to a microstructural dependent failure initiation [23]. Although significant improvements in the mechanical performance of printed parts has been achieved thus far, conventional HIP processes are unable to modify the columnar prior β-grains of AM processed parts [24]. Here we present a new approach that combines the deliberate introduction of a large density of lack of fusion defects (LoF) achieved via low laser energy density printing, and a single HIP treatment to simultaneously modify the microstructure and close pores in printed Ti – 6Al – 4V.

## 2. Materials and Methods

### 2.1. Selective laser melting of Ti – 6Al – 4V

The gas atomized Ti-6Al-4V powders (Carpenter Technology, Philadelphia, PA) with a particle size distribution between 10 – 45 μm was used in this study. Rectangular blocks of 32 mm in length, 6 mm width and 15 mm height were printed using the selective laser melting (SLM) process. Laser power and scanning velocities representing fully dense (FD) (300 W, 700 mm·s$^{-1}$) and lack of fusion (LoF) (100 W, 1300 mm·s$^{-1}$) regimes were chosen from microstructural analysis of a range of process parameters and analysis of their defect density (Figure 1). The additive process was carried out in the Open Additive printer (Beavercreek, OH) with an enclosed chamber and protective Argon atmosphere to prevent oxidation. A constant scanning strategy was implemented for all specimens consisting of a stripe pattern with a 67º rotation per layer. Layer height and hatch spacing were fixed at 50 μm for all printing conditions.

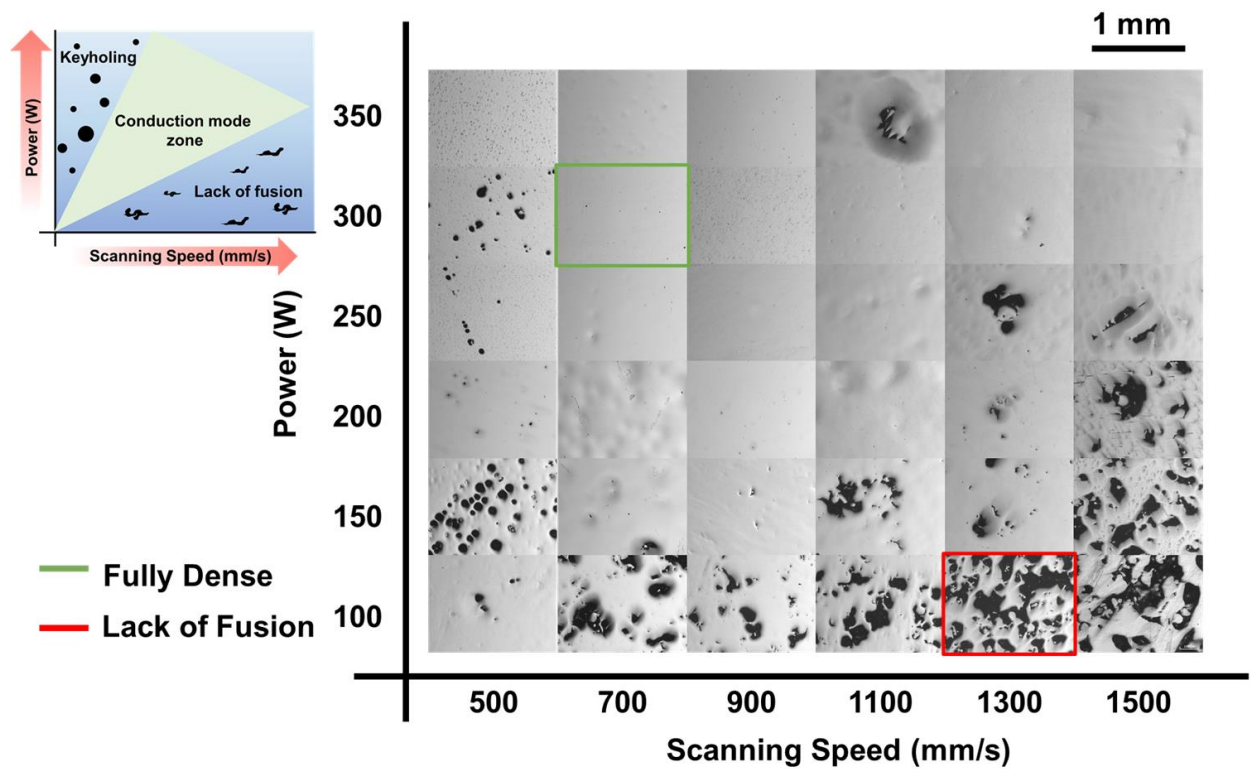

**Figure 1 Processing map of Ti-6Al-4V** Scanning speed vs. power variations in printing Ti-6Al-4V demonstrating three distinct regimes (keyholing, lack of fusion and conduction mode zones). Inset schematically shows the different processing regimes for Ti-6Al-4V. Highlighted micrographs represent the two sample groups used in this study to represent the FD condition (green), and LoF (red).

### 2.2. Heat treatment via hot isostatic pressing

As-printed specimens were subjected to a single HIP treatment using a QIH9 US HIP (Quintus Technologies, LLC, Lewis Center, OH) furnace in an Argon atmosphere. Heat treatments were carried out at a heating rate of 13°C·min$^{-1}$ using sub-transus (HIP 1) and super-transus (HIP 2) temperatures of 900°C and 1000°C for periods of 120 min and 60 min, respectively. All specimens were held at a maximum pressure of 100 MPa, and were furnace cooled (cooling rates of ~ 25-28°C·min$^{-1}$, Figure S1).

### 2.3. Microstructural characterization

As-printed and HIP-ed specimens were sectioned parallel to the build direction for microstructural evaluation. Surfaces were prepared following standard metallographic procedures (grinding and polishing up to 0.05 µm colloidal SiC) to remove surface imperfections. Polished surfaces were chemically etched using Kroll's reagent to reveal microstructural features. An Olympus BH-2 (Tokyo, Japan) optical microscope was used to assess the microstructure of as-printed and HIP-ed specimens and ImageJ (NIH, Maryland, USA) was implemented as a post-processing tool to quantify grain characteristics. At least 40 grain measurements were acquired to have statistically significant data. Similarly, at least 5 optical micrographs with lowest magnifications (1.4 ×1.0 mm$^2$) were utilized for porosity measurements in as-printed conditions. A Tescan Mira3 field emission scanning electron microscope (FE-SEM) equipped with a backscattering detector was used for imaging of polished and fracture surfaces. The morphology of grains, evolution of texture and grain orientation were evaluated via electron backscatter diffraction (EBSD). EBSD measurements were acquired using a QUANTAX EBSD (Bruker, Billerica, MA, USA) apparatus covering 105 µm × 105 µm area and processed by ATEX open source software [25]. To validate that the chosen scan area represents the bulk of the sample, quantification of the edge-to-edge distance between LoF defects was performed based on optical micrographs (1.4 × 1 mm$^2$) randomly obtained from the polished surfaces. Using ImageJ, at least 5 randomly oriented line profiles were extracted to compute the distribution of edge-to-edge defect distance. In a sample size of n=103 defects, a positively skewed distribution with mean edge-to-edge distance of 99.4 ± 85.9 µm between defects was found (Figure S3). A high standard deviation of ± 85.9 µm characterizes the wide range in the edge-to-edge distance distribution (2.5 – 396. 7 µm) of LoF defects . Due to the highly skewed nature of the distribution, the mode of the population (i.e. 40.8 µm) is better suited to describe the population as the most frequent edge-to-edge distance. In addition, these defects with areas between 2.2 µm$^2$ – 0.28 mm$^2$ were found to cover up to 42 ± 4.2 % of the bulk sample representing the probability ($P$= 0.42) of encountering a region with defects over the sample volume. Given this statistical analysis, the EBSD scan area of 105 µm × 105 µm was selected to encompass an area large enough to exceed the mode and mean of the edge-to-edge defect distance. This approach ensures the incidence of defect effect in the randomly selected scan area. Thus, it is concluded that the observed microstructural and texture evolution in LoF-HIP specimens is not an isolated occurrence and can be found randomly throughout the bulk of the sample.

The β→α'→α phase transformations were studied based on the Burgers relations $(110)_\beta//(0001)_\alpha$ and $[111]_\beta//[11\bar{2}0]_\alpha$, allowing the reconstruction of prior β-grains from the final α+β microstructure. The automatic reconstruction of parent grains from the EBSD scans of as-printed and HIP-ed specimens were performed using ARGPE software [26]

### 2.4. Mechanical characterization

Evaluation of the mechanical properties of the as-printed and HIP-ed specimens was carried out under tensile loads using a Deben MT 2000 micro-tensile stage equipped with a 2 kN load cell (Deben UK Ltd, Suffolk, UK). Micro-tensile specimens with a gauge length of 8 mm, width of 2 mm and thickness of 0.8 mm were machined via wire EDM perpendicular to the build direction. Tensile specimens were grinded down to 0.6 mm thickness and polished (0.05 µm colloidal SiC) to capture surface deformation features. Tensile experiments were performed in displacement control mode at an average strain rate of $1.3 \times 10^{-3}$ s$^{-1}$. Non-contact real time evolution of strains was captured by a digital image correlation software (GOM, Braunschweig, Germany) from recorded tensile displacements.

The strain hardening exponent ($n$) is computed as the slope of the logarithmic true stress-strain curves past the onset of plasticity following the power-law relationship:

$$\sigma = K\varepsilon^n \tag{1}$$

In which, $\sigma$ is the uniaxial true stress experienced by the specimen under tension, $K$ represents the strength coefficient, $\varepsilon$ is the true plastic strain and $n$ is the strain hardening exponent.

## 3. Results and Discussion
### 3.1. Microstructural evolution of AM Ti-6Al-4V

Laser power of P=300 W and scanning speed of V=700 mm·s$^{-1}$ leads to fully dense (FD) deposits of Ti-6A-4V with an average porosity of 0.6 ± 0.5 % (Figure 2 a). Five times reduction in energy density (P=100 W, V=1300 mm·s$^{-1}$) is implemented to intentionally introduce a high density of LoF defects (2.2 µm² – 0.28 mm²) resulting in an average porosity of 42% ± 4.2 %, as shown in (Figure 2b). The low energy density regime, also known as LoF zone, is unable to establish a melt pool with sufficient depths to bond with previously deposited layers resulting in fusion defects with areas ranging from 2.2 µm² – 0.28 mm²) (Figure 2b).

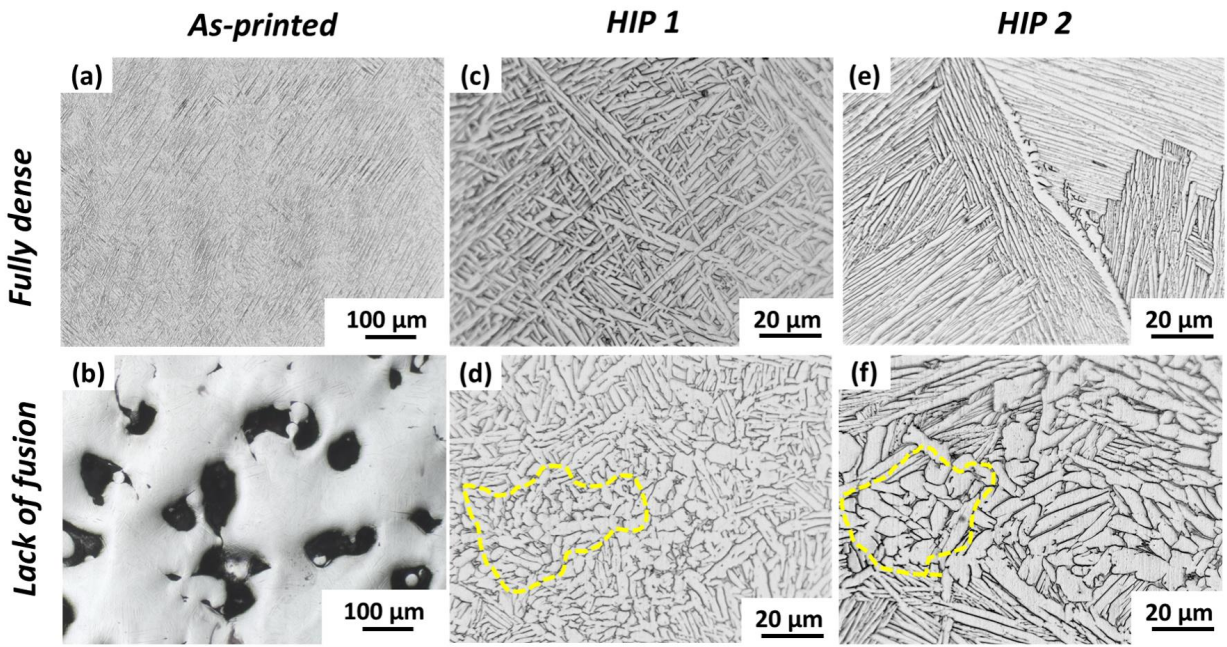

**Figure 2. Microstructural evolution of AM processed Ti alloy** Optical micrographs displaying the microstructure parallel to the building direction of as-printed specimens in (a) fully dense, (b)

lack of fusion printing regimes. Microstructural evolution of each printing regime as a result of sub-transus (c, d) and super-transus HIP treatments (E, F). Low aspect ratio α-grains emerged in specimens with a high density of lack of fusion (LoF) defects are highlighted to showcase the duplex microstructure attained after HIP 1 and HIP 2 treatments.

The microstructure of the as-printed specimens (Figure 2 a and b) is characterized by acicular grains, typical of the martensitic phase in Ti alloys [1,4,8,13]. Both LoF and FD samples are subjected to HIP with temperatures below (HIP 1 - 900 °C, 2 h) and above (HIP 2-1000 °C, 1 h) the transus temperature of the Ti-6Al-4V alloy ($T_β$=995 °C) [27]. After HIP, metastable α'-martensite transforms into α+β microstructure (Figure S2). The microstructure of the FD-HIP 1 samples is governed solely by Widmanstatten α-laths maintaining the morphology of α'-grains (Figure 2 c). On the other hand, HIP 1 condition led to the formation of a duplex microstructure in the LoF sample while fully eliminating the pores as shown in Figure 2 d. This microstructure has a large population of low aspect ratio grains with diameters ranging from 2.7-7.2 μm surrounded by elongated α-grains that can extend up to ~27 μm along their major axis. Post-processing of FD specimens using HIP 2 conditions (1000°C) resulted in the formation of α colonies with similar crystal orientation within prior β-grains (Figure 2 e). In the absence of LoF defects, the size of α-colonies grows to 183.2 ± 21.7 μm. Applying HIP 2 post-treatment to LoF samples resulted in the duplex microstructure similar to HIP 1 condition but with a much bigger grain size and colonies of 86.1 ± 8.9 μm, as shown in Figure 2 f.

Inverse Pole Figure (IPF) shows that the sub-transus HIP does not alter the strong crystallographic texture of the FD AM processed alloy (Figure 3 a and b) shown by maximum multiples of uniform density (MUD) values between 15-17.5.

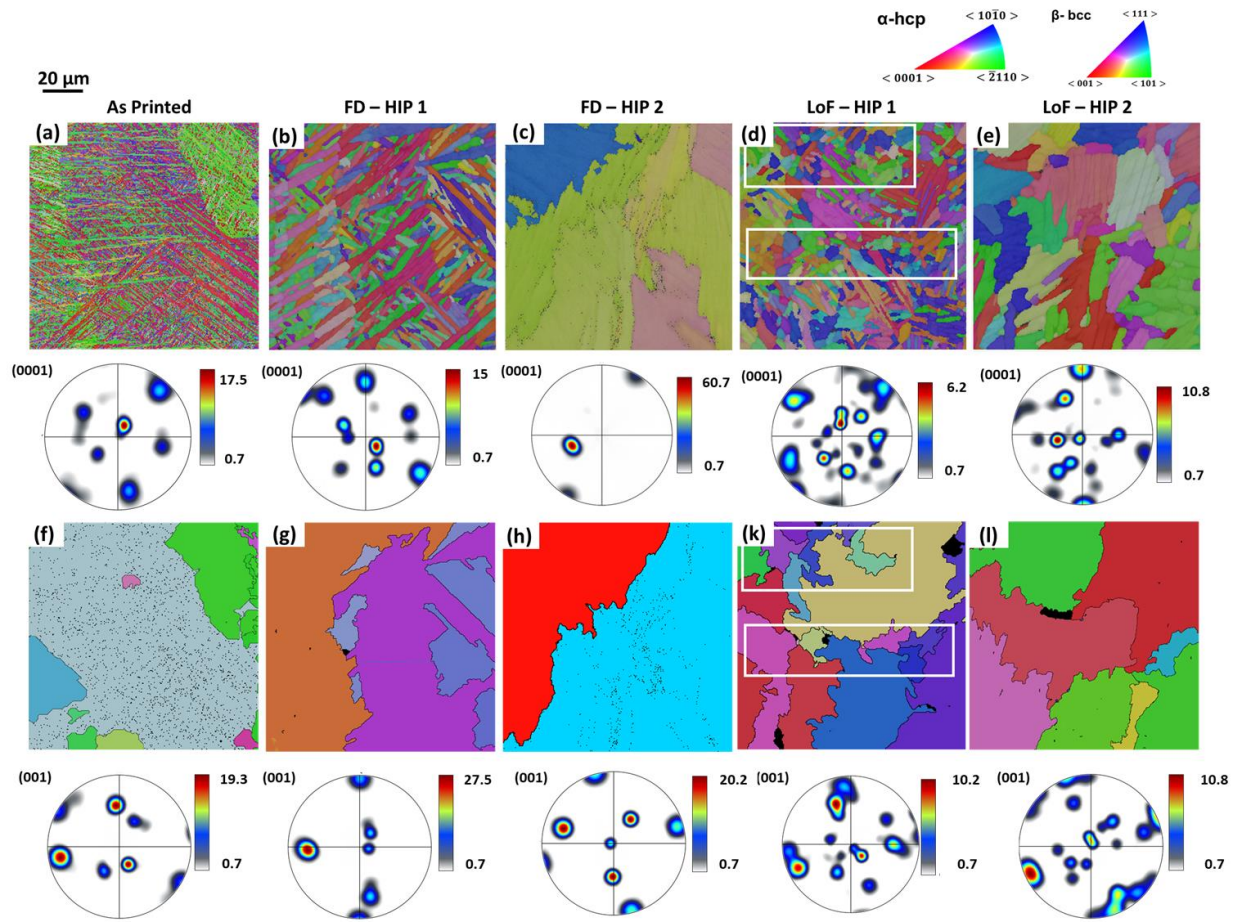

**Figure 3. Texture evolution in AM processed Ti-6Al-4V after HIP treatment** Inverse pole figure maps taken on the surface parallel to the building direction for the α-phase with respective pole figures along the (0001) pole of (a) FD as-printed, (b) FD-HIP 1 (c) FD-HIP 2, (d) LoF-HIP 1 and (e) LoF-HIP 2 specimens. (f-l) Recontructed prior β-grains and corresponding β-phase orientation maps obtained from the α-phase orientation maps (a-l).

Martensitic α'-grains and stable α-lamellae exhibit a preferred crystallographic orientation with their c-axis at ~ 20.5°-22° with respect to the (0001) pole. A slightly weaker texture in the FD-HIP 1 α-crystals (maximum MUD=9) with their c-axis tilted to 48.3° with respect to the (0001) pole is observed from the ease of slip across stable α/β interfaces enabling the slight rotation of the α-crystals [28–30]. In contrast, LoF-HIP 1 specimens (Figure 3 d) have a reduced maximum MUD of 6.2 as compared to as-printed FD (17.5) and FD-HIP 1 (15.0) specimens, showing a reduced texture in the duplex microstructure. Two weak texture components representing α-grains with their c-axis at ~20.3 ° and 45.1° (Figure 3 d) correspond to similar orientations of α-grains in the as-printed and FD-HIP 1 specimens. IPF maps confirm the formation of α-colonies in the alloy after HIP-ing above the $T_β$ (Figure 3 c and e) in both printing regimes with an intensified crystallographic texture.

These findings are in agreement with reconstructed prior β-grains maps (Figure 3 f-l), where reduced maximum MUD values of 10.2 (LoF-HIP 1) from 27.5 (FD-HIP 1) represent the ability to modify the microstructure and crystallographic texture of as-printed specimens exclusively in the LoF treated samples. As compared with all other samples, LoF-HIP 1 has smaller, irregularly shaped β-grains with reduced aspect ratios that interrupt the columnar grain growth (Figure 3 k). Regions with small prior β-grains in the β-reconstructed IPF maps outlined in white are directly associated with the group of equiaxed α-grains found in the corresponding α-phase orientation map (Figure 3 d). Given the close relationship between prior β-grains and the nucleation of α-grains, number density of prior β-grains was computed for all specimens [16]. Remarkably, the prior β-grain number density increases by more than 111% in LoF-HIP 1 ($17.1×10^{-4}$ $μm^{-2}$) as compared to FD-HIP 1 ($8.1×10^{-4}$ $μm^{-2}$) specimens. The effect of closed LoF defects on the microstructural changes is more pronounced in specimens treated at HIP 2 condition, where the β grain number density sees a five-fold increase in LoF-HIP 2 ($8.1×10^{-4}$ $μm^{-2}$) as compared to FD-HIP 2 ($1.7×10^{-4}$ $μm^{-2}$). In the absence of external nucleant particles, nucleation of the α-phase is known to commonly occur from the prior β-grain boundary[31]. Therefore, a higher number density of prior β-grains and the obvious decrease in size of prior β-grains is indicative of the increased nucleation sites for α-grains during HIP. This observation confirms that an initially high density of LoF defects in as-printed conditions led to the nucleation of new grains during HIP and is capable of modifying the microstructure of the titanium alloy. It is important to note, that although the texture analysis outlined in Figure 3 shows a localized scan area (105 μm × 105 μm), these are considered valid representations of the bulk of the specimen as shown via statistical analysis (section 2.3).

To support the significant difference in grain morphology observed in samples with an initial high density of LoF defects, a statistical significance analysis was performed between the population of grains in the FD-HIP 1 and LoF-HIP 1 specimens via a two-tailed z-test. Statistical examination of the difference in morphology of α-grains was evaluated based on their ellipticity ($E = 1 - (\frac{minor\ axis}{major\ axis})$), where grains with lower aspect ratio are described by their deviation from unity and values closer to zero. The populations of FD-HIP 1 and LoF-HIP 1 were considered to have normal distributions with sample sizes of n=625 and n=662, respectively. The absence of divergence among the ellipticity means of both populations was considered as the null hypothesis. A mean ellipticity of 0.50 in FD-HIP 1 specimens and 0.43 in LoF-HIP 1 was computed. Statistical analysis revealed a calculated z-test value of 5.6, allowing us to reject the null hypothesis with a 95% confidence level. This implies that the estimated probability of observing the decreased ellipticity in grains in the LoF-HIP 1 (Figure 3) by random chance is of $1.47 \times 10^{-8}$. Thus, the decreased aspect ratio in LoF-HIP 1 grains as a result of LoF defect closure after HIP has substantial statistical significance as compared to the grain ellipticity distribution of FD-HIP 1 specimens.

### 3.2. Role of lack of fusion defects in formation of duplex microstructure

The role of LoF defects in the formation of a duplex microstructure during HIP in the AM processed Ti alloy is discussed on the basis of two phenomena: (i) the reduction of high energy surfaces, and (ii) a local dislocation-driven recrystallization during the closure of LoF defects.

The healing of LoF defects by the thermo-mechanical HIP treatment is driven by a pressure assisted diffusion process [32]. The abundant free surfaces available in the LoF samples have excess surface energy making them ideal sites for the nucleation of new grains upon heating [33].

Heterogenous nucleation in a solid-state process can be described by the thermodynamic reduction of the activation energy barrier ($\Delta G_{het}$) described as [33,34]:

$$\Delta G_{het} = -V(\Delta G_V - \Delta G_s) + A\gamma - \Delta G_d \qquad (2)$$

Where V is the volume of the nuclei, $\Delta G_V$ and $\Delta G_s$ represent the change in free energy associated with a volume change and misfit strain energy respectively. $A$ is the area of the new interface with an associated interfacial energy $\gamma$, and $\Delta G_d$ is the excess free energy of a defect (i.e. free surfaces). It is clear from Eq. 2 that the elimination of a fusion defect (free surface) will result in the release of excess free energy ($\Delta G_d$), lowering the energy barrier for heterogenous nucleation of stress-free α-grains [33].

In addition to the reduction of surface energy, a secondary mechanism responsible for the nucleation of low aspect ratio α-grains is proposed to arise from the increase in stored energy ($E_{ss}$) available to drive the recovery and recrystallization process (Figure 4).

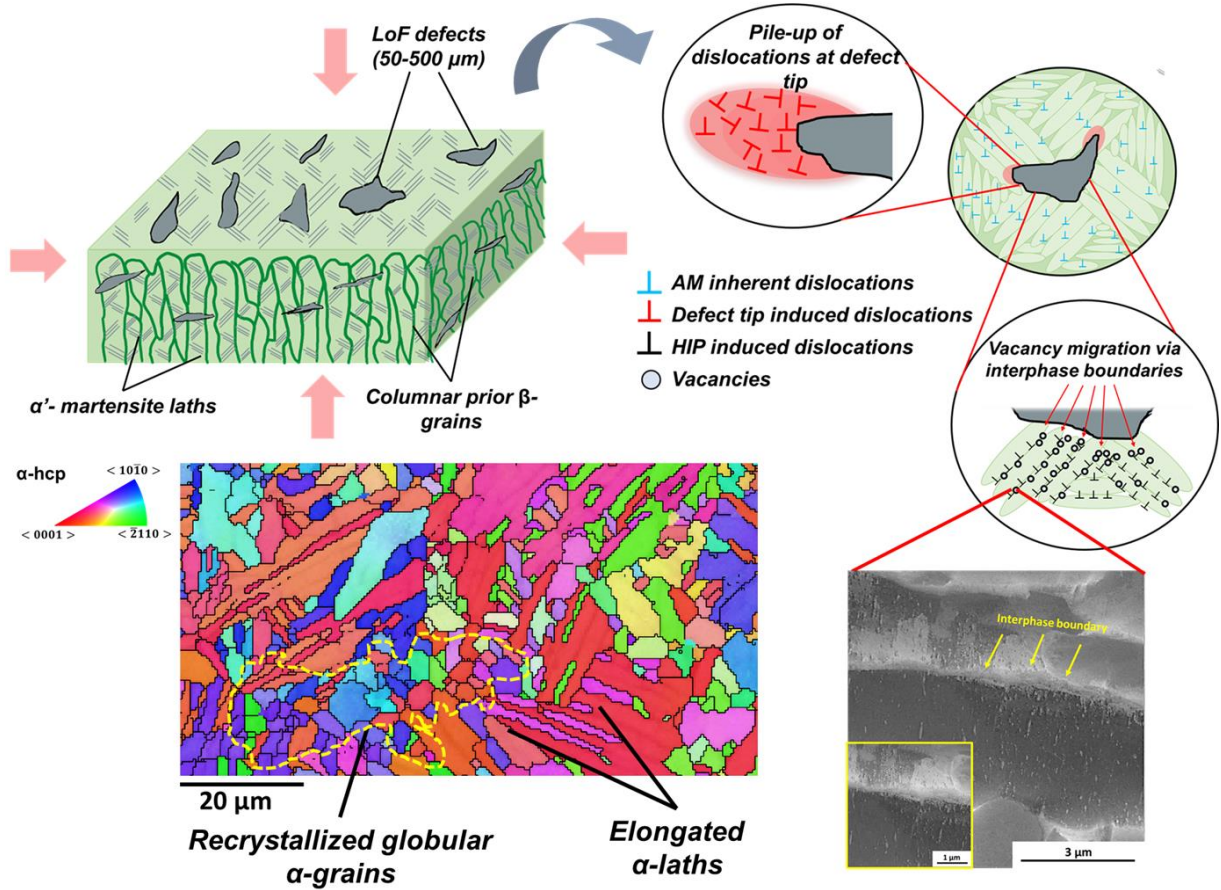

**Figure 4. Formation of duplex microstructure in AM processed Ti-6Al-4V** Schematic representation of the dislocation evolution in a LoF specimens during the early stages of HIP treatements: Generation of high density of dislocatiosn ahead of the LoF defect tips, vacancy migration via interphase boundaries and inherent dislocations derived from the martensitic phases in as-printed conditions. Resultant duplex microstructure of LoF-HIP 1 specimens after undergoing HIP promoting the nucleation and growth of low aspect ratio α-grains. Inset shows an electron channeling image (ECCI) depicting the pile-up of dislocations at the interphase between the α and β phase.

In the presence of LoF defects, $E_{ss}$ (Eq. 3) is proposed to have contributions from dislocations emerging ahead of the LoF defects ($E_{Kd}$), the accumulation of dislocations at the interphase boundaries (i.e. α/β-phase interfaces) due to vacancy diffusion ($E_{IB}$), and inherent strain energy ($E_s$) associated with the non-equilibrium martensitic microstructure [33]:

$$E_{ss} = E_{Kd} + E_{IB} + E_s \qquad (3)$$

Under the effect of high isostatic pressure in HIP, fusion defects with tip radius of 9 – 26 µm can be estimated to experience a high localized stress at the tip of the defect[35–37]. The localized high stresses are expected to be capable of emitting mobile dislocations from the tip of the LoF defect at the initial stage of the HIP process ($E_{Kd}$) [35–37]. With time, elevated temperatures will induce an *in-situ* recrystallization process leading to the nucleation of α-grains with reduced strain near prior LoF defects. A similar phenomenon has been reported in the healing of cracks in low carbon steel, nickel, and copper under the effect of thermo-mechanical treatments [34,38,39].

In addition, healing of pores via HIP process has been reported to occur via a combination of Nabarro-Herring and Coble creep mechanisms leading to the diffusion of vacancies [32,40]. Under the effect of hydrostatic stresses, the α/β interphases have been evidenced to serve as preferred areas for the pile-up of dislocations in response to HIP induced plastic deformations (Figure 4) [32]. These interphase boundaries with high dislocation densities are primary avenues to the healing of pores via vacancy diffusion mechanisms[32,41]. In this case, vacancies diffuse along dislocation cores pushing HIP-induced dislocations near the surface of the pores ($E_{IB}$) with diffusion rates 3-4 orders of magnitude higher than that of bulk diffusion[32,41]. The vacancy migration is further intensified in the present study due to the existence of higher density of porosities (up to 42 ± 4.2 %) increasing the stored lattice energy even further. The proposed dislocation-driven recrystallization process is more pronounced in LoF specimens due to the higher availability of stored strain energy as compared to the FD specimen with minimal porosity <0.6% where $E_{ss} \cong E_s$. Prior findings on the requirements for globularization in a Ti-6Al-4V via combined heating-deformation processes are in close agreement with the thermomechanical conditions experienced by LoF-HIP samples[28]. Prior studies have shown that strain induced dislocation processes can drive the globularization via recrystallization in an initially Widmanstätten microstructure[28,42,43]. To assert the proposed mechanism of localized recrystallization at LoF defect sites, the distribution of grain orientation spread (GOS) over the EBSD area is computed (Figure S4). Recrystallized regions can be characterized by a low value ($< 2°$) of GOS[42–44]. Quantification showed 50% and 30% of the scanned area in LoF – HIP 1 and FD-HIP 1 to be comprised of recrystallized grains, respectively. Given that the two specimens were subjected to the same HIP conditions, one can deduce that the 20% difference in recrystallized area corresponds to recrystallization events driven primarily by solid-state mechanisms that took place during the closure of LoF via HIP (surface energy reduction, dislocation emission at defect sites, and vacancy migration via interphase boundaries). A similar assertion where plastic deformation is the main driver for the microstructural changes near AM defects was reported by Li et al.[23]. Through a systematic analysis considering the length of the major axis in α-grains near defects, the authors confirmed localized grain globularization as a response to defect healing [23].

It is worth mentioning that the re-growth of gas-filled porosity has been reported in the past to occur under a super-transus heat treatment conditions (HIP at 1035 °C – 1200 °C) after their closure. Their regrowth has been attributed to the existence of Argon in pores [22,45]. Given that specimens produced in this study were manufactured via SLM in an Ar environment, LoF defects could be filled by Ar. Thus, closed defects could potentially experience re-growth in the form of spherical pores after further heat treatments. It is noteworthy that re-growth of LoF defects is absent in samples printed by electron beam melting due to the lack of an inert gas during the printing process (i.e. printing conducted in vacuum) [22,45].

In this work, the proposed mechanism, and observations of distinct microstructural evolution in LoF-HIP 1 and LoF-HIP 2 conditions are corroborated via EBSD orientation maps and statistical analysis of grain morphology. This study proves the potential to engineer microstructures via the introduction of LoF defects and their subsequent closure via thermo-mechanical treatments.

### 3.3. Mechanical properties and mechanisms of deformation

The mechanical performance of the designed microstructures was evaluated under tensile loading perpendicular to the building direction. These directions are known to exhibit a reduced tensile ductility due to the inherited anisotropy of the additive process[1]. Tensile experiments were performed on as-printed and HIP-ed specimens (Figure 5 a), and their mechanical properties are summarized in Table S1.

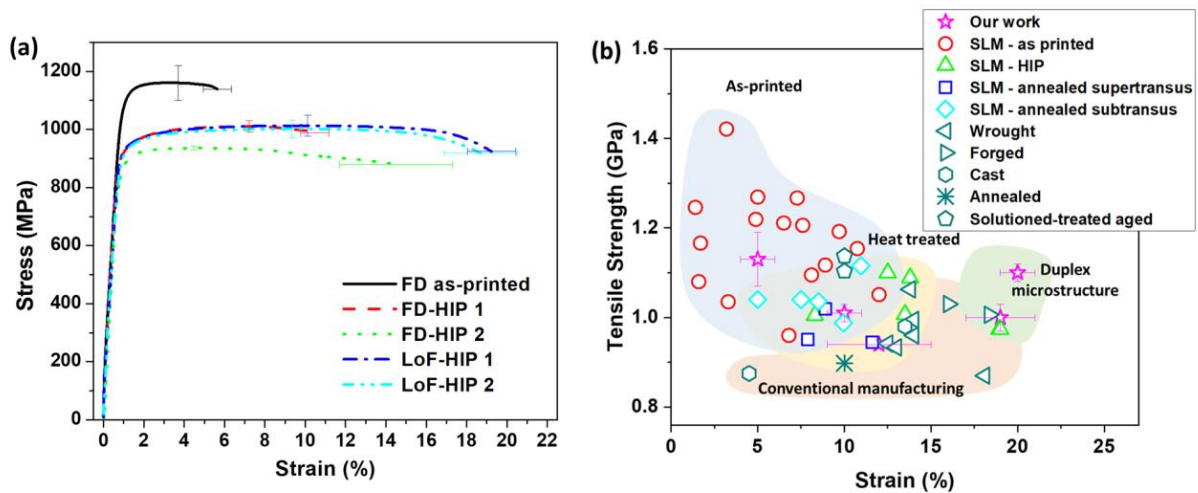

**Figure 5. Mechanical properties of as-printed and HIP Ti-6Al-4V.** (a)Average stress-strain response curves of as-printed and HIP-ed specimens under tensile loading. (b)Ashby plot showing a significant improvement in tensile strength and failure strain shown by the engineered duplex microstructures (LoF-HIP 1 and HIP 2) vs. reported conventional processes (wrought, forged, cast, annealed, and solutioned-treated aged), as-printed and HIP-ed AM Ti alloys[4].

All heat-treated specimens experience a decrease in their yield strength associated with the reduction of dislocation density in the microstructure, stabilization of phases and grain growth. This effect is more pronounced in FD-HIP 2 where easy slip transmission across laths with similar crystal orientation occurs (Figure 6 a and b), while all the other heat-treated samples retain a comparable yield strength. In contrast, LoF-HIP 2 shows a similar yield strength to the samples heat treated at sub-transus temperatures. The retained strength is attributed to the smaller α-colony size of 86.1 ± 8.9 µm as compared to FD-HIP 2 (183.2 ± 21.7 µm), and the presence of low aspect ratio α grains with random crystal orientations to effectively confine the slip length (Figure 7 c and d). More interestingly, the LoF-HIP 1 and HIP 2 specimens are able to retain their tensile strength (1.0 ± 0.02 GPa) while showing an unprecedented increase in failure strain by up to 300% and 90% as compared to FD as-printed (0.05 ± 0.01) and FD-HIP 1 (0.10 ±0.01), respectively. To put the obtained results into context, Figure 5 b compares the tensile strength-ductility of our work with other data available in the literature on SLM and heat treated SLM (SLM-HT), cast and wrought Ti-6Al-4V[4]. It can be observed that the introduction of LoF-induced duplex microstructures is capable of achieving an unprecedented ductility with comparable strength to heat treated SLM data extending the property space of AM Ti alloys. Even more, LoF-HIP 1 and HIP 2 outperform conventional wrought, forged, cast, annealed, and solutioned-treated and aged processes that are considered as benchmark to compare the properties of Ti-6Al-4V processed via AM[27].

The role of α-colonies on the deformation behavior of samples HIP-ed at super-transus temperature is shown in Figure 6 a and b. The post-failure surfaces show the formation of slip traces in a single direction within α-colonies (183.2 ± 21.7 µm), and their intersection at the α-grain boundaries (corresponding to prior β grain boundaries) where there is a change in crystal orientation (Figure 6 a and b).

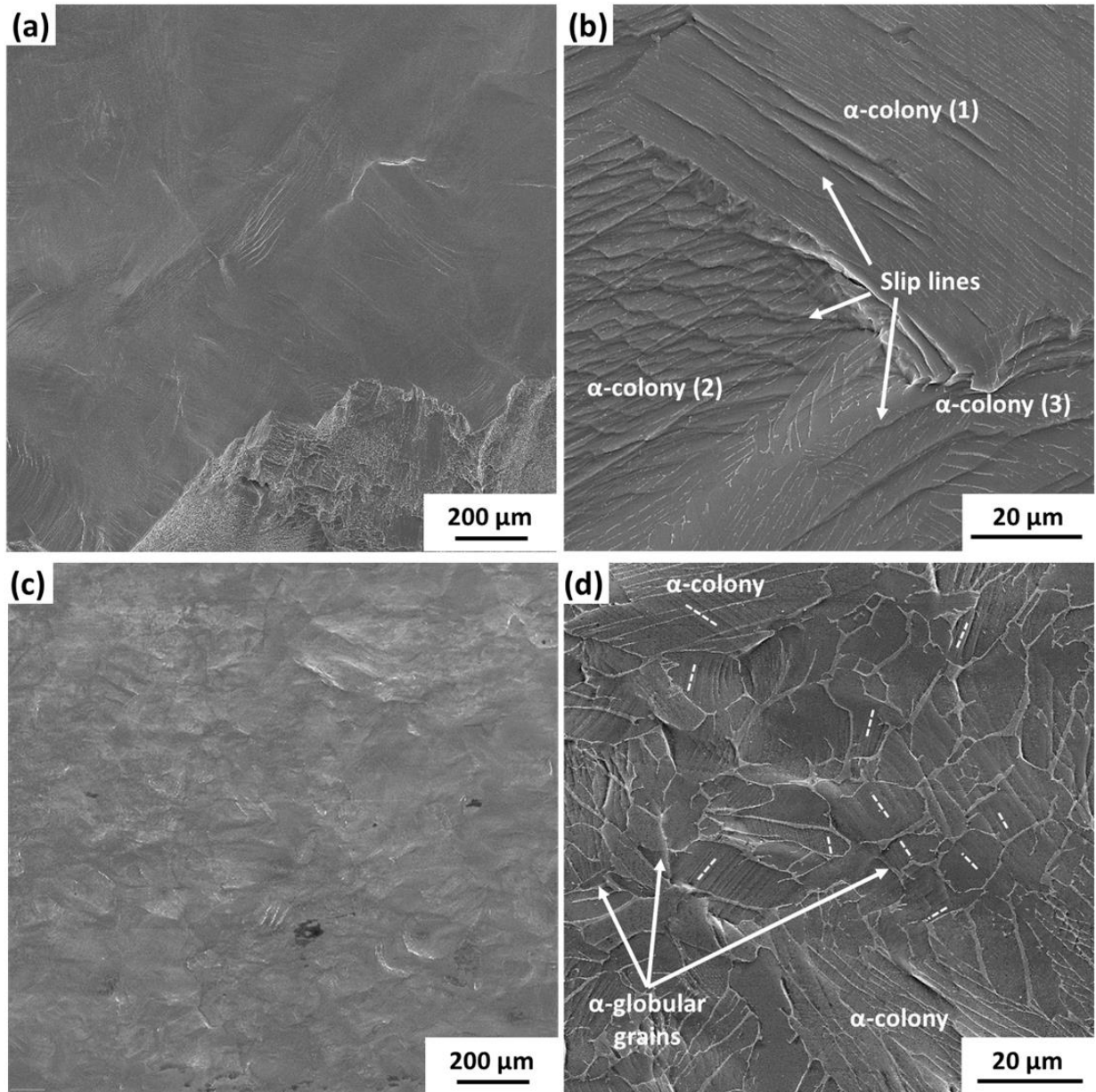

**Figure 6. Role of α-colonies on deformation behavior** Post-mortem investigation of surface steps of FD-HIP 2 specimens showing (a) the macroscale (b) detailed view of deformation governed by formation of slip traces predominantly in a single direction within a single α-colony and change of slip direction at the intersection of α colonies with different crystal orientations; (c) LoF-HIP 2 showing macroscale deformation, (d) slip traces along a single direction across multiple α-laths in colonies, intersecting low aspect ratio α- grains with slip traces in random orientations.

Analysis of the fracture surface and surface steps reveals the evolution of deformation mode from a brittle fracture in the as-printed samples to ductile fracture in the LoF-HIP 1 samples. As a result, an increase in dimple features on the fracture surface (Figure S3) and non-directional surface steps (Figure 7a-c) can be observed.

The plastic deformation in FD-HIP 1 specimens is mainly limited to plasticity at the interphase boundaries as discerned by the steep surface steps originating at the α/β interphase boundaries in the surfaces parallel to the TD (Figure 7 b). On the other hand, the non-directional surface steps in LoF-HIP 1 are indicative of the extended deformation enabled by the presence of low aspect ratio grains with no preferential crystallographic orientation (Figure 7 c).

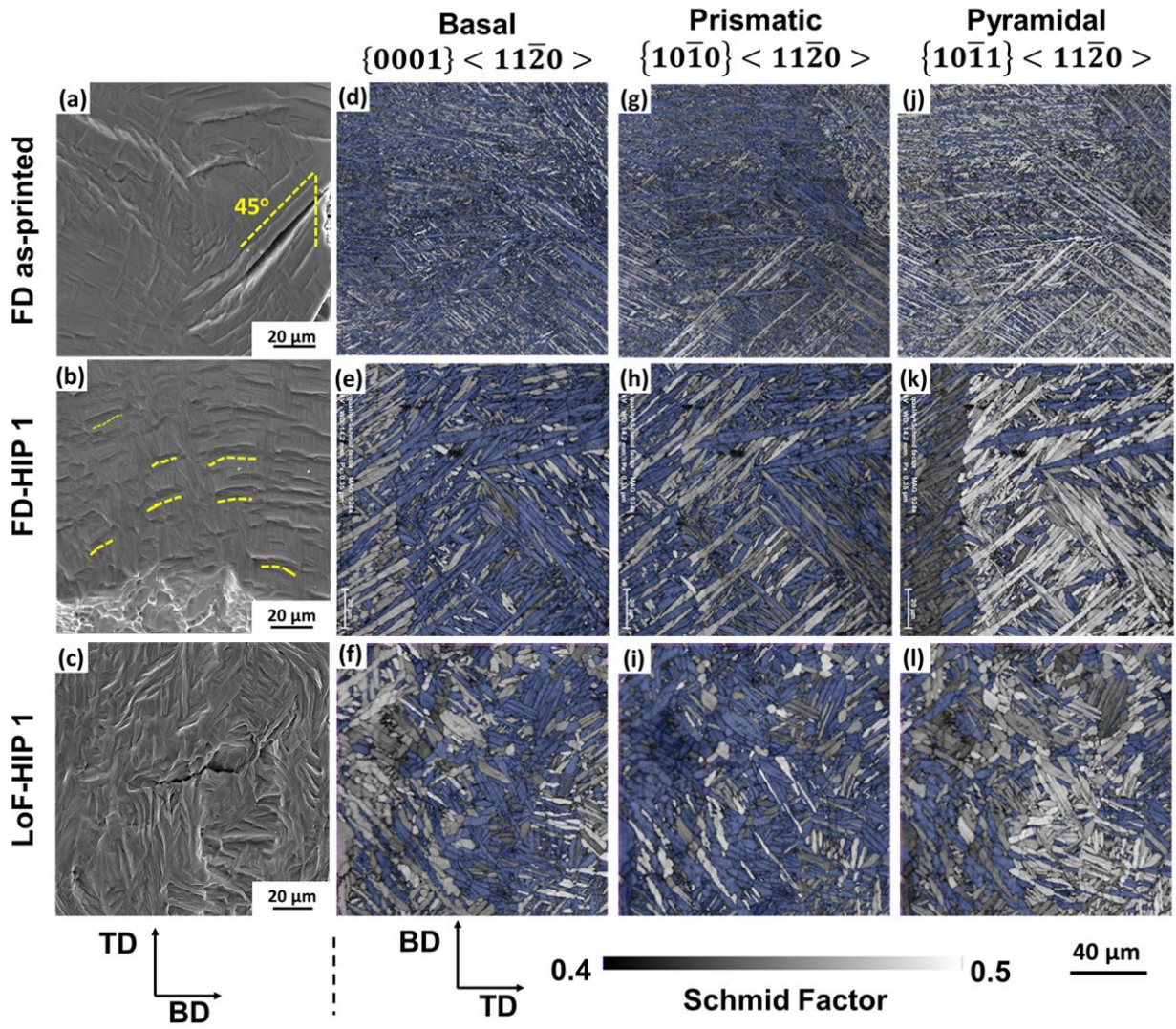

**Figure 7. Deformation mechanism of AM fabricated Ti-6Al-4V.** Post-mortem surface characterization of (a) as-printed specimens showing shearing-type deformation initiating at primary α'-laths. (b) Limited plasticity in FD-HIP 1 specimens characterized by radial surface steps at the α/β interphase boundaries. (c) Extended plasticity in the LoF-HIP 1 specimens can be observed by high density of non-directional surface steps. (d-l) Schmid factor maps of basal, prismatic, and pyramidal slip systems. Grains with a Schmid factor of 0.4-0.5 are shown in gray-scale.

To understand the mechanism by which these materials extend their ductility without significantly compromising strength, we study the distribution of potential active slip systems in grains with high Schmid factor (SF) (Figure 7 d-l). Computation of SF distribution was performed considering the tensile axis is perpendicular to the build direction (x-axis). The image quality + SF distribution maps shown in Figure 7 d-l display grains with high SF of 0.4 – 0.5 in a gray-scale (representing crystals with an orientation suitable for the activation of the corresponding slip system), while those with SF <0.4 are shown in blue. Area fractions of basal, prismatic, and pyramidal slip systems in the as-printed microstructures are 47%, 55%, and 73%, respectively. A similar distribution is determined in FD-HIP 1, with area fractions of 44% basal, 53% prismatic and 78% pyramidal. In contrast, LoF-HIP 1 shows an increase in the area fraction of grains with a preferred basal slip of 62%, including a large population of the newly nucleated α- grains with low aspect ratios. A comparable distribution corresponding to prismatic slip of 43% and 73% pyramidal is attributed to the retained textured of α-laths. Basal and prismatic slip systems are main activated slip systems at room temperature in the α-Ti phase due to their low critical resolved shear stress (CRSS) [46]. The higher volume of α-grains with orientations suitable for the activation of basal slip in LoF-HIP 1 specimens, suggest that the emerged low aspect ratio α-grains contribute to the extended plasticity (Figure 7 f). Despite the FD-HIP 1 and LoF-HIP 1 exhibiting a similar capacity to work-harden (Table S1), the higher fraction of α-grains with orientations favoring basal and prismatic slip in LoF-HIP 1 enable a gradual strain redistribution introducing a quasi-stable stress flow regime that is absent in as-printed and FD-HIP 1 specimens (Figure 6 a)[47]. The considerable degree of surface damage sustained by the specimen and dimple structure on the fracture surface (Figure 7 c and Figure S5) evidences the presence of a ductile fracture behavior as compared to its counterparts. These findings imply the ability to prevent immediate fracture of AM Ti alloys by controlling grain orientation and morphology to extend their ductility while preserving strength.

It is important to note that the remarkable mechanical performance in the LoF-HIP 1 specimen results from a single thermo-mechanical post process without the inclusion of foreign particles and/or changes in the alloy composition.

## 4. Conclusion

This study moves away from the established intuition that LoF defects should always be avoided and demonstrates the possibility to exploit these process inherent defects, in addition to standardized HIP process, as a pathway to print alloys with tailored microstructures and enhanced mechanical properties. This novel processing pathway results in emergence of a *duplex microstructure* that is revealed to be driven by a dislocation-induced recrystallization and reduction of surface energy intensified by the presence of LoF defects. The *duplex microstructure* exhibits unprecedented plasticity as compared to as-printed, HIP-ed, forged, annealed, solution-treated and aged, cast and wrought Ti-6Al-4V without compromising its strength. It is important to highlight that the technique to engineer duplex microstructures presented herein is not restricted to the SLM process. Rather, it can be applied to various AM processes to spatially distribute LoF defects by controlling the energy density during printing. In addition, while the present work focuses on Ti-6Al-4V, the same principles can be applied to a wide-range of metals and alloys suffering from undesirable columnar microstructures in AM including steels, Ni based superalloys [48–50] and Aluminum alloys [51,52].

**Data and materials availability**

The datasets generated during and/or analyzed during the current study are available from the corresponding author on reasonable request.

**Declaration of competing interests**

The authors declare that they have no known competing financial interests or personal relationships that could have appeared to influence the work reported in this paper.

**Authorship contribution statement:**

A.M contributed to the conceptualization, writing-review & editing of the scientific work, J.B. contributed to the investigation, writing – review & editing, methodology, and data curation. J.K. contributed to the investigation, review & editing, methodology. All authors discussed the results and edited the manuscript at all stages.

**Acknowledgments**

Jenniffer Bustillos gratefully acknowledges funding received by the 2020-2021 Knight @ KIC Engineering Graduate Fellowship. The authors acknowledge Open Additive, LLC (Beavercreek, OH) and John Middendorf for the fabrication of SLM specimens used in this study. We also acknowledge Magnus Ahlfors at Quintus Technologies, LLC (Lewis Center, OH) for services provided in HIP post-processing of specimens. This work made use of the Cornell Center for Materials Research Shared Facilities which are supported through the NSF MRSEC program (DMR-1719875).

**Supplementary Materials:**

Supplementary material associated with this article can be found in the online version

42. Kim, I.-S. *et al.* Accelerating globularization in additively manufactured Ti-6Al-4V by exploiting martensitic laths. *J. Mater. Res. Technol.* **12**, 304–315 (2021).

43. Huang, K. & Logé, R. E. A review of dynamic recrystallization phenomena in metallic materials. *Mater. Des.* **111**, 548–574 (2016).

44. Cao, Y. *et al.* An electron backscattered diffraction study on the dynamic recrystallization behavior of a nickel-chromium alloy (800H) during hot deformation. *Mater. Sci. Eng. A* **585**, 71–85 (2013).

45. Tammas-Williams, S., Withers, P. J., Todd, I. & Prangnell, P. B. Porosity regrowth during heat treatment of hot isostatically pressed additively manufactured titanium components. *Scr. Mater.* **122**, 72–76 (2016).

46. Bridier, F., Villechaise, P. & Mendez, J. Analysis of the different slip systems activated by tension in a α/β titanium alloy in relation with local crystallographic orientation. *Acta Mater.* **53**, 555–567 (2005).

47. Hwang, J. K. Revealing the small post-necking elongation in twinning-induced plasticity steels. *J. Mater. Sci.* **55**, 8285–8302 (2020).

48. Balachandramurthi, A. R. *et al.* Microstructure tailoring in Electron Beam Powder Bed Fusion additive manufacturing and its potential consequences. *Results Mater.* **1**, 100017 (2019).

49. Helmer, H., Bauereiß, A., Singer, R. F. & Körner, C. Grain structure evolution in Inconel 718 during selective electron beam melting. *Mater. Sci. Eng. A* **668**, 180–187 (2016).

50. Bean, G. E. *et al.* Build Orientation Effects on Texture and Mechanical Properties of Selective Laser Melting Inconel 718. *J. Mater. Eng. Perform.* **28**, 1942–1949 (2019).

51. Thijs, L., Kempen, K., Kruth, J. P. & Van Humbeeck, J. Fine-structured aluminium products with controllable texture by selective laser melting of pre-alloyed AlSi10Mg powder. *Acta Mater.* **61**, 1809–1819 (2013).

52. Martin, J. H. *et al.* 3D printing of high-strength aluminium alloys. *Nature* **549**, 365–369 (2017).

# Supplementary File

## Lack of fusion in additive manufacturing: defect or asset?

*Jenniffer Bustillos[1], Jinyeon Kim[1], Atieh Moridi[1*]*

[1]*Sibley School of Mechanical and Aerospace Engineering*
*Cornell University*
*124 Hoy Road, 469 Upson Hall*
*Ithaca, NY 14853, USA*
[*]*Corresponding author:moridi@cornell.edu*

**Table S1** Mechanical properties and computed strain hardening exponent of the as-printed and HIPed specimens.

| Sample | Elastic Modulus (GPa) | Yield Strength (MPa) | Ultimate Tensile Strength (GPa) | Failure strain (%) | Strain hardening exponent (*n*) |
|---|---|---|---|---|---|
| FD - as printed | 119.1 ± 0.8 | 1095.4 ± 56.4 | 1.1 ±6E-2 | 5 ± 1 | 0.03 ± 2.3E-4 |
| FD - HIP 1 | 123.8 ± 1.3 | 945.9 ± 30 | 1.01 ± 2E-2 | 10 ±1 | 0.07 ± 1.2E-4 |
| FD - HIP 2 | 117.6 ± 3.6 | 908.2 ± 10.4 | 0.9 ± 1E-2 | 12 ± 3 | 0.05 ± 0.8E-4 |
| LoF – HIP 1 | 108.16 ± 6.8 | 925.5 ± 21 | 1.0 ± 3E-2 | 20 ± 1 | 0.08 ± 2.1E-4 |
| LoF – HIP 2 | 112.1± 7.9 | 930.0 ± 30.3 | 1.0 ± 3E-2 | 19 ± 2 | 0.07 ± 1.8E-4 |

## Supplementary Discussion

### Thermo-mechanical treatment of AM Ti-6Al-4V

Figure S1 shows the heating curves used in the hot isostatic pressing (HIP) of the AM Ti-6Al-4V. Using a heating rate of 13°C/min, samples were heat treated by implementing maximum temperatures of 900 °C (sub-transus, HIP 1) and 1000°C (super-transus, HIP 2).. All experiments were carried out in an Argon atmosphere, held at a maximum pressure of 100 MPa and were furnace cooled.

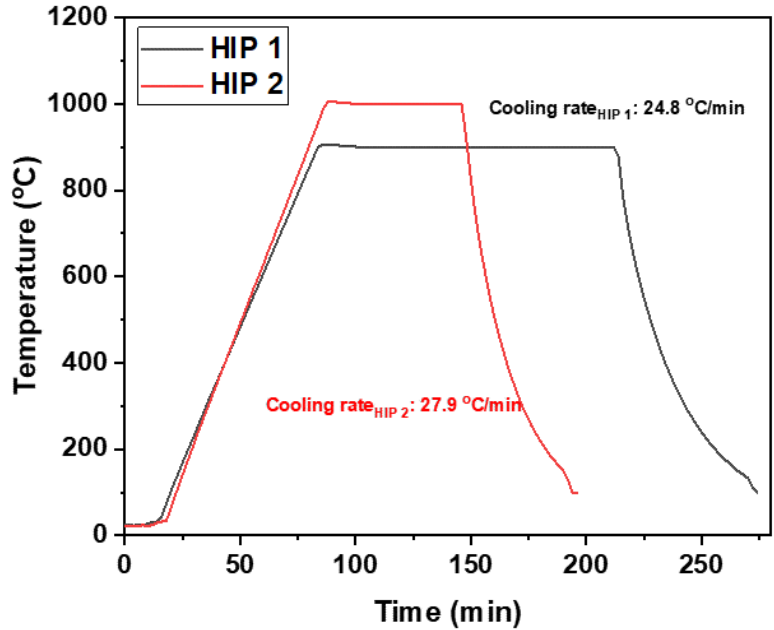

**Figure S1** Representative heating curves of the thermomechanical HIP treatment carried out at subtransus (HIP 1) and supertransus (HIP 2) temperatures.

### Microstructural evolution of AM Ti-6Al-4V

Figure S2 shows the backscatter electron micrograph image of as-printed fully dense (FD) and lack of fusion (LoF) specimens before and after undergoing sub- and super-transus HIP. FD as-printed microstructure is characterized by fine acicular α' martensite corresponding to non-equilibrium phases developed during the rapid solidification of additive manufacturing (Figure S2a).

At sub-transus HIP (HIP 1, 900°C, 100 MPa, 2 h), α'-martensite stabilizes and results in α-laths (shown in dark contrast) with β-phase at the grain boundaries (light contrast) as shown in Figure S2b. Higher temperatures past the β-transus temperature of the Ti-6Al-4V alloy (HIP 2, 1000°C, 100 MPa, 1 h) results in the nucleation and growth of elongated α-laths sharing a single crystal orientation in each colony. Colonies are shown to be separated via α-grain boundaries, known to be detrimental to the mechanical performance of the alloy[1]. The introduction of LoF defects in the as-printed microstructure results in the nucleation and growth of strain free α-grains with globular morphology via recrystallization after HIP 1 schedule (See enclosed regions in Figure S2 d, e). Super-transus HIP of LoF specimens retains the duplex microstructure and growth of α-colonies is restricted by newly nucleated grains around prior defect surfaces (Figure S2 f).

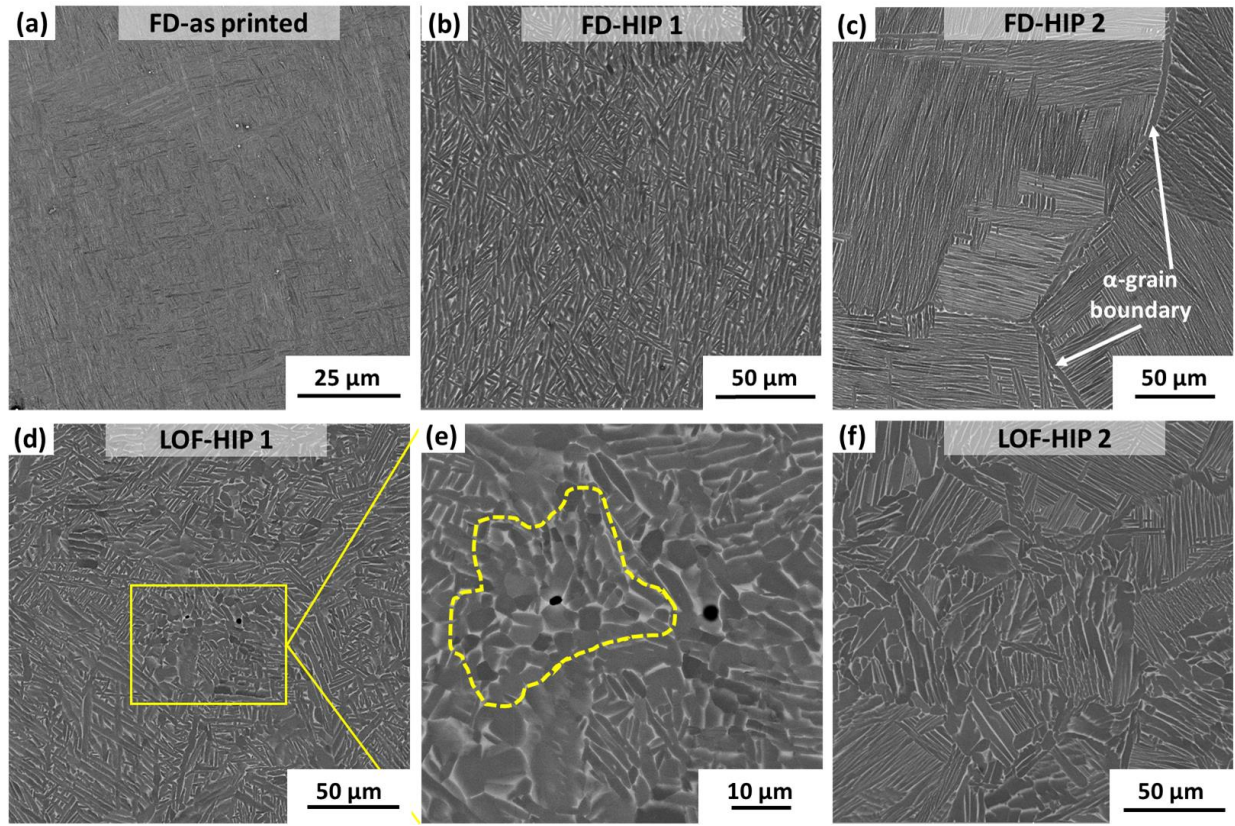

**Figure S2 Microstructural and phase characteristics of AM and HIP-ed Ti6Al4V**
Backscatter electron images demonstrating distribution of (a) fine acicular α' phases in FD as-printed, (b) stabilized α-phase (darker contrast) and β-phase residing at the grain boundaries (lighter contrast) in FD-HIP 1, (c) α-phases colonies in the form of elongated laths surrounded by α-grain boundaries, (d,e) duplex microstructure of the LoF-HIP 1 showing recrystallized globular α-grains around LoF defects surrounded by lamellar α-laths retained from the AM process, (f) restricted growth of α-colonies in super-transus HIP LoF specimens (LoF-HIP2).

**Selection of Electron Backscattering Diffraction (EBSD) scan area based on defect quantification methods**

Scanning area for EBSD was chosen based on the most recurring edge-to-edge distance between defects. The edge-to-edge distance between LoF defects was computed from randomly obtained optical micrographs with areas of 1.4 × 1 mm$^2$ of the LoF (as-printed) samples. Line profiles (5) from each image were used to extract a distribution of edge-to-edge defect distance shown in Figure S3. In a sample size of n=103 defects, the data set showed a positively skewed distribution (Figure S3) with mean edge-to-edge distance of 99.4 ± 85.9 μm. The high standard deviation (± 85.9 μm) characterizes the wide range of distances in the distribution (2.5 – 396.7 μm). Due to the highly skewed nature of the distribution, the mode of the population is computed to be of 40.8 μm, representing the most frequent edge-to-edge defect distance. Thus, a scan area of 105 μm × 105 μm was chosen to exceed both the mean and mode of the edge-to-edge defect distance. This approach will ensure that the randomly selected area will reflect the effect of LoF defects.

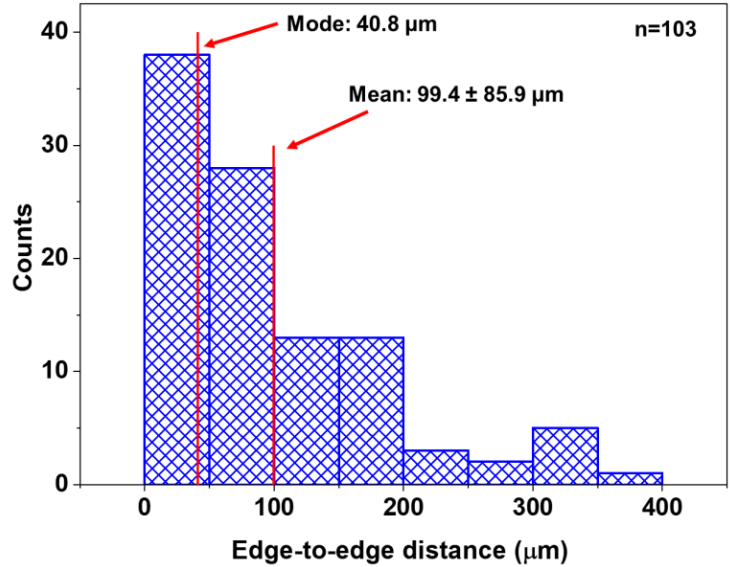

**Figure S3** Distribution of edge-to-edge defect distance in the as-printed LoF samples showing a skewed behavior with respective statistical mode and mean.

**Grain orientation spread distribution as a measure of recrystallization**

In an effort to isolate and quantify the recrystallization driven by the proposed LoF-HIP process, the grain orientation spread (GOS) distribution over the EBSD scanned area is computed (Figure S4). A GOS value of <2º was chosen to represent recrystallized grains as it represents the low dislocation density and internal misorientation of these grains. Figure S4 shows the distribution of GOS up to 5º in grains within the scanned EBSD areas for FD-HIP 1 and LoF-HIP 1 samples.

FD-HIP 1 specimens showed 30% of the area to correspond to recrystallized grains, while LoF-HIP 1 showed a much higher area of 50% representing recrystallized grains. It is inferred that the difference in area with low GOS values (<2º) is attributed to the effect of LoF defect-induced recrystallization during HIP.

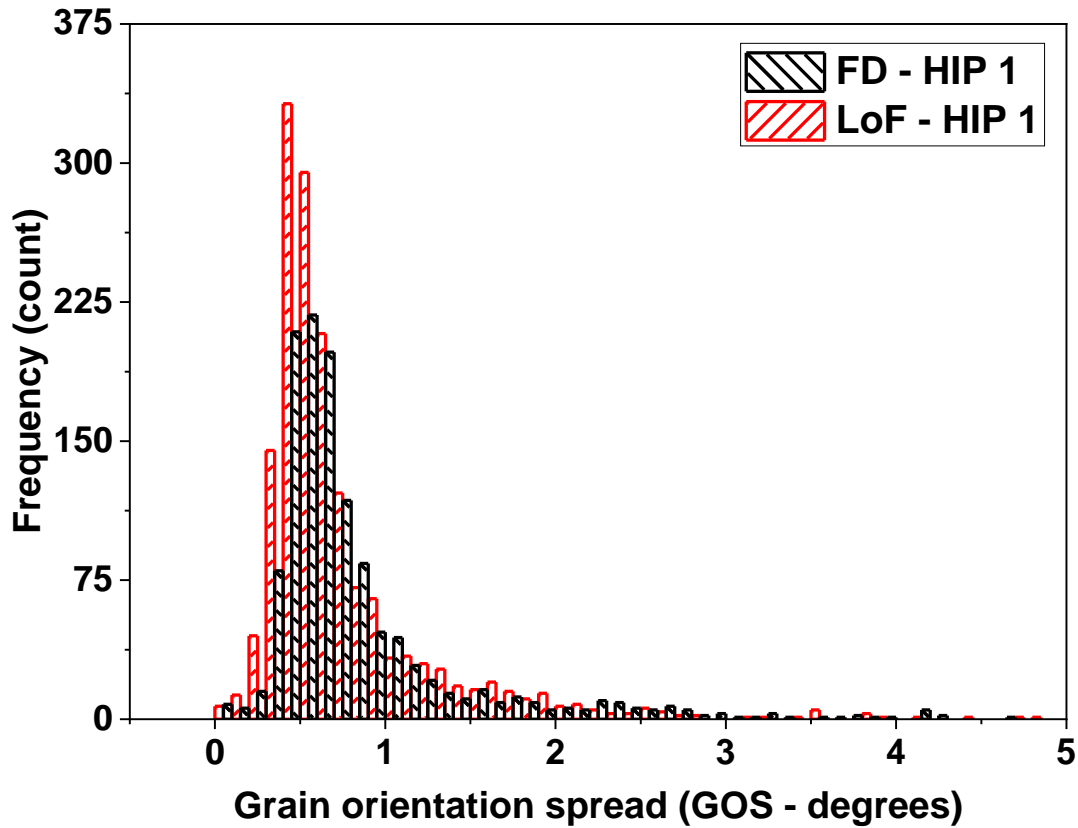

**Figure S4** EBSD grain orientation spread distribution of LoF-HIP 1 and FD-HIP 1 specimens. A low GOS of < 2º is used to characterize recrystallized grains with low dislocation densities.

**Fracture surface analysis of AM processed Ti-6Al-4V after HIP**

Evaluation of the fracture surface perpendicular to the loading direction reveals the presence of cleavage facets in the as-printed specimens indicative of a brittle failure at the macroscale (Figure S5a). The preferred orientation of the α' laths along the plane of maximum shear stress is responsible for the localization of strains in this region leading to the nucleation of voids and premature failure. This agrees with a prior study by our group, where in-situ tensile studies with concurrent micro digital image correlation (DIC) revealed the localization of strains in large aspect ratio primary α' laths[2]. These α' laths exhibited high Schmidt factors in the basal and prismatic slip systems suggesting their easy activation upon tensile loading leading to the nucleation of voids, their coalescence and final failure.

Stabilization of martensitic phases in the FD samples subjected to sub-transus HIP resulted in a slight change in the texture of the α-crystals (c-axis rotation from 22º to 48º). These microstructural and crystallographic changes enabled a limited enhancement in the failure strain ($\varepsilon_f$=10%). This effect can be evidenced in the fracture surface of the FD-HIP 1 specimen showing cleavage facets with limited dimple features on the surface (Figure S5 b).

On the other hand, fracture surfaces corresponding to the LoF-HIP 1 specimen (Figure S5c) show dimple features throughout the surface indicative of a ductile failure. This ductile deformation is attributed to the extended plasticity enabled by the presence of new low aspect ratio α-grains with orientations suitable for the activation of basal and prismatic slip systems.

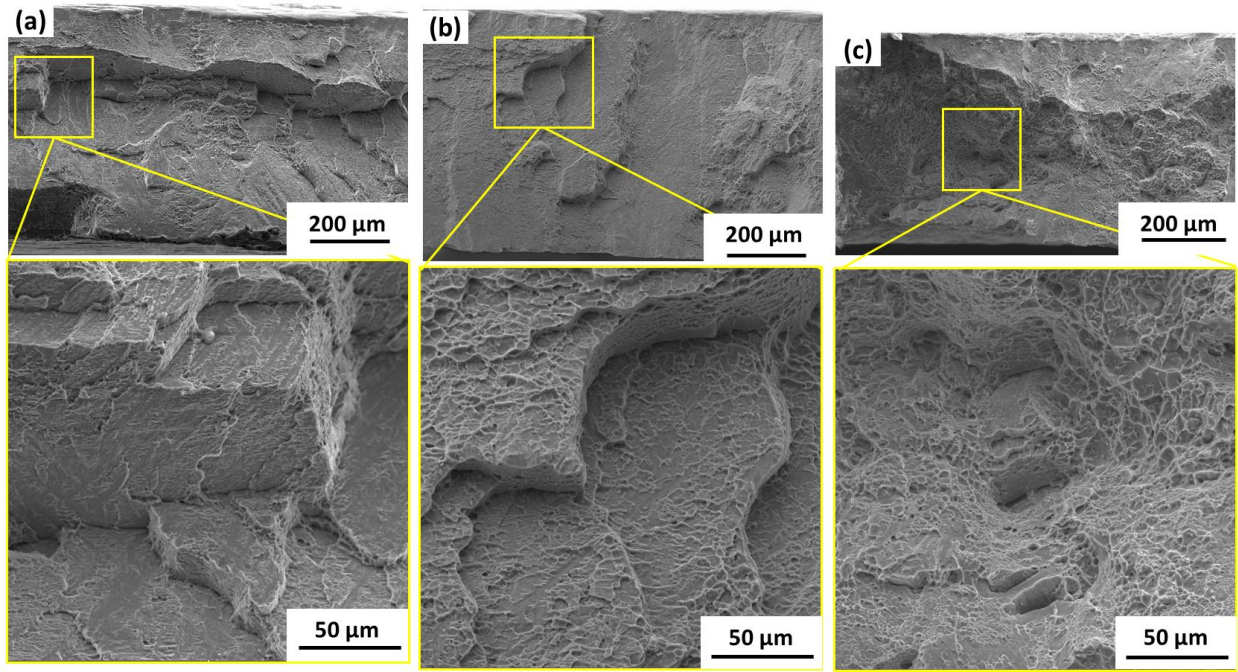

**Figure S5 Fracture surface in AM and HIP-ed Ti-6Al-4V** Scanning electron micrographs of the fracture surfaces perpendicular to the tensile load direction of (a) FD as-printed specimens showing sharp cleavage features characteristic of macroscopic brittle fracture, detailed view shows sharp steps with featureless surface, (b) FD-HIP 1 with planar fracture surface, detailed view shows the limited plasticity encountered after sub-transus HIP revealed by limited dimple like features, (c) LoF-HIP 1 shows irregular features in the reduced cross-sectional area due to extensive plasticity and necking achieved prior to fracture, detailed view shows a combination of dimple features and some clean facets characteristic of the ductile fracture in the duplex microstructure.